\documentclass[11pt]{article}

\title{Correlation Functions of Finite XXZ Model with Boundaries}
\author{Yasuhiro Fujii and Miki Wadati}
\date{Department of Physics, Graduate School of Science, \\
  University of Tokyo,\\  Hongo 7-3-1, Bunkyo-ku, Tokyo 113, Japan}

\renewcommand{\theequation}{\arabic{section}.\arabic{equation}}

\begin{document}
\maketitle

\begin{abstract}
  The finite XXZ model with boundaries is considered.
  We use the Matrix Product Ansatz (MPA),
  which was originally developed in the studies on
  the asymmetric simple exclusion process
  and the quantum antiferromagnetic spin chain.
  The MPA tells that the eigenstate of the Hamiltonian is constructed by
  the Zamolodchikov-Faddeev algebra (ZF-algebra) and the boundary states.
  We adopt the type I vertex operator of $U_q(\widehat{sl}_2)$
  as the ZF-algebra
  and realize the boundary states
  in the bosonic $U_q(\widehat{sl}_2)$ form.
  The correlation functions are given by the product of the vertex operators
  and the bosonic boundary states.
  We express them in the integration forms.
\end{abstract}

\setcounter{equation}{0}
\section{Introduction}
\label{sec:1}

Even for the integrable systems,
to calculate the $N$-point correlation functions is a very difficult problem.
In the XXZ model, the correlation functions are found
for the infinite \cite{CBMS,Nakayashiki,Foda}
and the half-infinite \cite{Jimbo,Miwa} chains
by the approach of the vertex operators \cite{Jing,Frenkel}.
Those vertex operators are constructed
in the bosonic $U_q(\widehat{sl}_2)$ form
and are known to have two types, type I and type II\@.
The former means the half-infinite product of the R-matrices,
and the latter describes the excitation of the model.
In the infinite case, it is natural that the quantum affine algebra
$U_q(\widehat{sl}_2)$ is adopted,
because the model has its symmetry, i.e.,
the Hamiltonian commutes with its generators.
However, the half-infinite XXZ model obviously has no such symmetries.
In analysis of that model,
it is important to know the physical meaning
of the type I vertex operator.
The transfer matrix can be defined by the type I vertex operator
without physical conflicts;
the boundary states are realized in terms of $U_q(\widehat{sl}_2)$
and the correlation functions are calculated by using its expression.
Since the vertex operators do not indicate the finite lattices physically,
in the finite chain with boundaries things go wrong in such a way.

Without regard to calculation of the correlation functions,
the analysis of the models with boundaries is in general more difficult
than that of the models without boundaries.
Roughly speaking, there are three theories to treat the boundary.
The first uses the boundary K-matrix
in the quantum integrable systems \cite{Cherednik,Sklyanin,Kulish}.
The K-matrix describes the integrable boundary conditions
and satisfies the reflection equation (see eq. (\ref{eqn:RE})).
Several solutions are reported \cite{Vega,Inami,Hou,Yung}.
The second is the boundary S-matrix method
in the integrable field theories \cite{Zamolodchikov,Faddeev,Ghoshal,Fendley}.
In such field theories, for instance, the conformal field theory,
the scattering matrix of particles is factorized
into a product of two-particle S-matrices,
and the boundary scattering is described
by the boundary S-matrix.
Those S-matrices are known to have similar properties
as the R-matrix and the K-matrix.

The third, called the Matrix Product Ansatz (MPA),
is different from the previous two methods.
It was originally developed in the studies on
the asymmetric simple exclusion process \cite{DEHP,Sandow,Rittenberg}
and the quantum antiferromagnetic spin
chain \cite{Klumper,Shadschneider,Zittarz}.
The asymmetric simple exclusion process is a stochastic process
where particles diffuse asymmetrically,
and are injected and ejected at boundaries.
Under the MPA,
the state on each site is represented by the infinite-dimensional matrices
determined by the model, and 
the correlation functions are defined by the products of those matrices.
Calculating them is formally possible.
However it is generally difficult
to treat such infinite-dimensional expressions.
For removing this difficulty,
the MPA has been recently extended by using the terminology
of the integrable systems \cite{Sasamoto}.
It has been proved that the infinite-dimensional matrices
correspond to the representations of the Zamolodchikov-Faddeev algebra
(ZF-algebra).
That is an algebra where the generators are ``scattered''
mutually by the R-matrix and at the boundary by the K-matrix.
In addition it has been shown that
a certain eigenstate of the Hamiltonian is given
by the product of the ZF-algebra and the boundary states.
The remarkable point is that the symmetry of the model is not required
in spite of using the ZF-algebra.
This property is the mirror of the original one.
The ZF-algebra and the boundary states are realized by comparing the result
with the known one in other theory.
By using them,
the eigenstates can be computed
irrespective of the symmetry of the model.
It is also possible to find the correlation functions.
The MPA thus offers a new powerful tool to analyze the model with boundaries.

The aim of this paper is to calculate the correlation functions
of the finite XXZ model with boundaries by applying the MPA\@.
We note that the type I vertex operators can be regarded as the ZF-algebra,
because they satisfy the same commutation relations.
We adopt the type I vertex operator as the ZF-algebra,
which is justified by the agreement of the result
and the known one in the half-infinite chain limit.
While the vertex operators indicate the half-infinite products
of the R-matrices,
such physical meanings are hidden
and only those mathematical properties are observed.
The reason is that the vertex operators are used
only for realization of the eigenstate through the ZF-algebra
and are not related with the symmetry of the model.
The detail is discussed in Section \ref{sec:2}.
We realize the boundary states
in the bosonic $U_q(\widehat{sl}_2)$ form
in Section \ref{sec:3}.
Based on those bosonic realization,
we calculate the $2N$-point function that generates the correlation functions
in Section \ref{sec:4}.
In Section \ref{sec:5} we discuss our results
and note some open problems.
The technical details are summarized
in Appendices \ref{sec:app1}, \ref{sec:app2} and \ref{sec:app3}.

\setcounter{equation}{0}
\section{Matrix Product Ansatz}
\label{sec:2}

\subsection{Formulation of Matrix Product Ansatz}

In this section we formulate the MPA by use of the theory
of the integrable systems \cite{Sasamoto}.
We consider the model with boundaries.
Let $R(\zeta)$ be the R-matrix and $K(\zeta)$ be the boundary K-matrix.
They satisfy the reflection equation \cite{Cherednik,Sklyanin},
\begin{equation}
\label{eqn:RE}
  K_2(\zeta_2)R_{21}(\zeta_1 \zeta_2)K_1(\zeta_1)R_{12}(\zeta_1/\zeta_2) =
  R_{21}(\zeta_1/\zeta_2)K_1(\zeta_1)R_{12}(\zeta_1 \zeta_2)K_2(\zeta_2).
\end{equation}
In customary $\zeta$ is called as the spectral parameter.
The lower labels attached to operators indicate
the lattice sites where they act.
The transfer matrix is defined by
\begin{equation}
  T(\zeta_N,\ldots,\zeta_1) =
  K_N^+(\zeta_N)
  \left(
    \prod_{i=1}^{\leftharpoonup \atop{N-1}}
    R_{i,N}(\zeta_i \zeta_N)
  \right)
  K_N^-(\zeta_N)
  \left(
    \prod_{i=1}^{\rightharpoonup \atop{N-1}}
    R_{N,i}(\zeta_N/\zeta_i)
  \right).
\end{equation}
The superscripts $+$ and $-$ of the K-matrices refer to
the left (site $N$) and right (site $1$) boundaries.
The ZF-algebra is a set of operators $F(\zeta)$ and $F^*(\zeta)$ satisfying
the following relations \cite{Zamolodchikov,Faddeev,Ghoshal,Sasamoto}.
\begin{eqnarray}
\label{eqn:def_ZF}
  R_{12}(\zeta_1/\zeta_2)F_1(\zeta_1)\otimes F_2(\zeta_2)
  &=&
  F_2(\zeta_2)\otimes F_1(\zeta_1), \\
\label{eqn:def_dualZF}
  R_{12}(\zeta_1/\zeta_2)F_1^*(\zeta_1)\otimes F_2^*(\zeta_2)
  &=&
  F_2^*(\zeta_2)\otimes F_1^*(\zeta_1), \\
\label{eqn:def_bo}
  \mathcal{V}\cdot F(\zeta^{-1})K^+(\zeta)
  &=&
  \mathcal{V}\cdot F(\zeta), \\
  K^-(\zeta)F(\zeta)\cdot \mathcal{W}
  &=&
  F(\zeta^{-1})\cdot \mathcal{W}, \\
  K^+(\zeta)F^*(\zeta)\cdot\mathcal{V^*}
  &=&
  F^*(\zeta^{-1})\cdot\mathcal{V^*}, \\
\label{eqn:def_dual_bo}
  \mathcal{W^*}\cdot F^*(\zeta^{-1})K^-(\zeta)
  &=&
  \mathcal{W^*}\cdot F^*(\zeta).
\end{eqnarray}
Here $\mathcal{V}$ and $\mathcal{W}$ are the left and right boundary states,
and $\mathcal{V}^*$ and $\mathcal{W}^*$ are those dual states.
The MPA claims that a product of the ZF-algebra
and the boundary states gives the eigenstate of the transfer matrix
whose eigenvalue is $1$.
Namely, the ``state''
\begin{eqnarray}
\label{eqn:state}
  &&
  |\zeta_N,\ldots,\zeta_1\rangle
  =
  |\mathcal{V}\cdot F_N(\zeta_N)\otimes\cdots\otimes F_1(\zeta_1)
  \cdot\mathcal{W}\rangle, \\
\label{eqn:dual_state}
  &&
  \langle\zeta_1,\ldots,\zeta_N|
  =
  \langle\mathcal{W}^*\cdot F_1^*(\zeta_1)\otimes\cdots\otimes F_N^*(\zeta_N)
  \cdot\mathcal{V}^*|
\end{eqnarray}
is the eigenstate of the transfer matrix.
We make the ansatz that the states (\ref{eqn:state}) and (\ref{eqn:dual_state})
have mathematical meanings.
The proof is straightforward.
Using the relations (\ref{eqn:def_ZF})-(\ref{eqn:def_dual_bo}), we have
\begin{eqnarray}
\label{eqn:prf_state}
  &&
  T(\zeta_N,\ldots,\zeta_1)
  |\zeta_N,\ldots,\zeta_1\rangle
  \nonumber \\
  && \quad =
  K_N^+(\zeta_N)
  \left(
    \prod_{i=1}^{\leftharpoonup \atop{N-1}}
    R_{i,N}(\zeta_i \zeta_N)
  \right)
  K_N^-(\zeta_N)
  |\zeta_{N-1},\ldots,\zeta_1,\zeta_N\rangle
  \nonumber \\
  && \quad =
  K_N^+(\zeta_N)
  \left(
    \prod_{i=1}^{\leftharpoonup \atop{N-1}}
    R_{i,N}(\zeta_i \zeta_N)
  \right)
  |\zeta_{N-1},\ldots,\zeta_1,\zeta_{N}^{-1}\rangle
  \nonumber \\
  && \quad =
  K_N^+(\zeta_N)
  |\zeta_{N}^{-1},\ldots,\zeta_1\rangle
  \nonumber \\
  && \quad =
  |\zeta_N,\ldots,\zeta_1\rangle.
\end{eqnarray}
Similarly, we can show that
\begin{equation}
  \langle\zeta_1,\ldots,\zeta_N|T(\zeta_N,\ldots,\zeta_1) =
  \langle\zeta_1,\ldots,\zeta_N|.
\end{equation}
Observing the relations
\begin{eqnarray}
  &&
  T(1,\ldots,1) = 1, \\
  && \hspace{-0.5cm}
  \frac{\partial}{\partial\zeta_N}
  T(\zeta_N,\ldots,\zeta_1)\biggr|_{\zeta_N=\cdots=\zeta_1=1} =
  \mbox{const.}\times H,
\end{eqnarray}
we further have
\begin{equation}
  H|1,\ldots,1\rangle = 0,
  \qquad
  \langle1,\ldots,1|H = 0.
\end{equation}
We conventionally call the states (\ref{eqn:state}) and (\ref{eqn:dual_state})
as the stationary state and the dual stationary state.
In particular cases the stationary state corresponds to the ground state.
For instance, in the half-infinite XXZ model,
the stationary state is known to be the ground state
when $|r|<1$ \cite{Jimbo}.
(See eq. (\ref{eqn:delta,r}) for the definition of the parameter $r$.)
The stationary state is one of the candidates for the ground state.

Note that the proof (\ref{eqn:prf_state}) is similar to
the derivation of the equations that the correlation functions satisfy
in the half-infinite XXZ model \cite{Miwa}.
The difference between them is the definition of the boundary states.
While in the usual theory
the boundary states mean the eigenstate of the Hamiltonian,
in the MPA they indicate the boundaries themselves.
The eigenstate in the MPA depends on the ZF-algebra and the boundary states,
and the model is defined by the R-matrix and the K-matrix.
It is important to notice that the boundaries and the Hamiltonian
are treated independently.

We indicate that the realization of the ZF-algebra is
independent of the number of the sites of the model.
In the MPA the ZF-algebra is defined by an interaction on the nearest sites
and the number of the sites of the model is given by
the number of the ZF-algebras appeared
in the stationary state (see (\ref{eqn:state})).
Then if the ZF-algebra is realized for a particular number of the sites,
such realization is always correct.

\subsection{Application to Finite XXZ Model with Boundaries}

Let us apply the MPA to the finite XXZ model with boundaries.
The Hamiltonian of the model is
\begin{equation}
\label{eqn:H}
  H =
  -\frac{1}{2} \sum_{i=1}^{N-1}
  (\sigma_{i+1}^{x}\sigma_1^x
  +\sigma_{i+1}^{y}\sigma_1^y
  +\Delta\sigma_{i+1}^{z}\sigma_1^z)
  +h^+\sigma_N^z
  +h^-\sigma_1^z,
\end{equation}
where $\sigma_i^\alpha$ ($\alpha=x,y,z$) are the Pauli matrices acting
on a site $i$,
\begin{equation}
  \sigma_i^x=\left[
    \begin{array}{rr}
      0 & 1 \\
      1 & 0
    \end{array}
  \right]_i,
  \quad
  \sigma_i^y=\left[
    \begin{array}{rr}
      0 & -\mbox{i} \\
      \mbox{i} & 0
    \end{array}
  \right]_i,
  \quad
  \sigma_i^z=\left[
    \begin{array}{rr}
      1 & 0 \\
      0 & -1
    \end{array}
  \right]_i.
\end{equation}
The R-matrix and the K-matrix are defined in Appendix \ref{sec:app1}.
The derivative of the transfer matrix gives the Hamiltonian,
\begin{equation}
  \frac{\partial}{\partial\zeta_N}
  T(\zeta_N,\ldots,\zeta_1)\biggr|_{\zeta_N=\cdots=\zeta_1=1}
  =
  \frac{4q}{1-q^2}H.
\end{equation}
The constants $\Delta$ and $h^\pm$ are parameterized as
\begin{equation}
\label{eqn:delta,r}
  \Delta = \frac{q+q^{-1}}{2},
  \qquad
  h^\pm = -\frac{1-q^2}{4q}\frac{1+r_\pm}{1-r_\pm}.
\end{equation}
We consider the model in the massive regime $\Delta<-1$, i.e., $-1<q<0$.
In the MPA, one must realize the ZF-algebra and the boundary states
satisfying the relations (\ref{eqn:def_ZF})-(\ref{eqn:def_dual_bo}).
We adopt the type I vertex operator $\Phi(\zeta)$ of $U_q(\widehat{sl}_2)$
as the ZF-algebra
and assume that the boundary states can be described
in terms of $U_q(\widehat{sl}_2)$.
(The vertex operators of $U_q(\widehat{sl}_2)$ are formulated
in Appendix \ref{sec:app2}.)
By the definition (\ref{eqn:state}) the stationary state is defined by
\begin{equation}
\label{eqn:def_state}
  |\zeta_N,\ldots,\zeta_1\rangle
  =
  \sqrt{G}\times
  |\mathcal{V}\cdot
  \Phi(\zeta_N)\otimes\cdots\otimes\Phi(\zeta_1)
  \cdot \mathcal{W}\rangle,
\end{equation}
with some normalization factor $G$.
By using the dual vertex operators $\Phi^*(\zeta)$,
the dual stationary state is similarly defined by
\begin{equation}
\label{eqn:def_dual}
  \langle\zeta_1,\ldots,\zeta_N|
  =
  \sqrt{G}\times
  \langle \mathcal{W}^*\cdot
  \Phi^*(\zeta_1)\otimes\cdots\otimes\Phi^*(\zeta_N)
  \cdot \mathcal{V}^*|.
\end{equation}
With these definitions
we express the correlation function involving the local operator $\mathcal{O}$
by
\begin{eqnarray}
\label{eqn:def_corr}
  &&
  \langle\zeta_1,\ldots,\zeta_N|
  \mathcal{O}
  |\zeta_N,\ldots,\zeta_1\rangle
  \nonumber \\
  && \quad =
  G\times
  \langle \mathcal{W}^*\cdot
  \Phi^*(\zeta_1)\otimes\cdots\otimes\Phi^*(\zeta_N)
  \cdot \mathcal{V}^*|
  \mathcal{O}|\mathcal{V}\cdot
  \Phi(\zeta_N)\otimes\cdots\otimes\Phi(\zeta_1)
  \cdot \mathcal{W}\rangle.
\end{eqnarray}
The definitions (\ref{eqn:def_state}) and (\ref{eqn:def_dual})
are justified by  the verification of the expression (\ref{eqn:def_corr})
in the limit $N\rightarrow\infty$.
We assume that the local operator $\mathcal{O}$ has a $2\times 2$ matrix form
$\mathcal{O}=\sum_{m=1}^{N}(\mathcal{O}_m)_{\epsilon\epsilon'}$.
In the limit $N\rightarrow\infty$
the boundary states $\mathcal{V}$ and $\mathcal{V}^*$ are ignored.
Applying the invertibility of the vertex operators (\ref{eqn:inv})
we find that
\begin{eqnarray}
\label{eqn:N-inf}
  &&
  \langle\zeta_1,\ldots,\zeta_N|
  \mathcal{O}
  |\zeta_N,\ldots,\zeta_1\rangle
  \nonumber \\
  && \quad =
  g^{m-N}G\times
  \sum_{m=1}^{N}\sum_{\epsilon_1,\ldots,\epsilon_m,\epsilon'_m}
  \langle \mathcal{W}^*\cdot
  \Phi_{\epsilon_1}^{*}(\zeta_1)\cdots\Phi_{\epsilon_{m-1}}^{*}(\zeta_{m-1})
  \Phi_{\epsilon_m}^{*}(\zeta_m)
  \nonumber \\
  && \qquad \times
  (\mathcal{O}_m)_{\epsilon_m \epsilon'_m}
  \Phi_{\epsilon'_m}(\zeta_m)
  \Phi_{\epsilon_{m-1}}(\zeta_{m-1})\cdots\Phi_{\epsilon_1}(\zeta_1)
  \cdot \mathcal{W}\rangle,
\end{eqnarray}
where $g$ is a scaler constant given in eq. (\ref{eqn:g}).
We have used the vector expressions of the vertex operators
$\Phi(\zeta)$ and $\Phi^*(\zeta)$
(see eqs. (\ref{eqn:def_ver_map}) and (\ref{eqn:def_d_ver_map})),
\begin{equation}
  \Phi(\zeta)=\sum_{\epsilon=\pm}\Phi_\epsilon(\zeta)\otimes v_\epsilon,
  \qquad
  \Phi^*(\zeta)=\sum_{\epsilon=\pm}\Phi_\epsilon^*(\zeta)\otimes v_\epsilon^*.
\end{equation}
We set $G=g^N/\langle\mathcal{W}^*\cdot\mbox{id}|\mbox{id}\cdot\mathcal{W}\rangle$.
If $\langle\mathcal{W}^*\cdot\mbox{id}|$ and $|\mbox{id}\cdot\mathcal{W}\rangle$
are the same as the eigenstates of the half-infinite XXZ model,
the correlation function (\ref{eqn:N-inf})
is identical to the already known result \cite{Jimbo}.
(Notice that our boundary states correspond to the eigenstates
in the usual theories.)
In the next section we shall compute the boundary states
and make sure that this conjecture is indeed correct.
The correlation functions correspond to the known results
in the limit $N\rightarrow\infty$.
Note that the realization of the ZF-algebra is independent of
the number of the sites of the model.
The justification is thus approved.

Remarks are in order.
First, in the present case, $U_q(\widehat{sl}_2)$ does not play
the role of a symmetry,
because it does not commute with the Hamiltonian (\ref{eqn:H}).
We adopt the vertex operators for realization of the MPA
and disregard the mathematical symmetry of the model.
It is not requested in the MPA\@.
Second, one might consider that the physical meanings of
the vertex operators conflict to finiteness of the model.
This doubt is reversed by the following reasons.
In the usual theory the vertex operators constitute the transfer matrix
and prescribe the model through it, however,
in the MPA they realize the ZF-algebra and give the eigenstate
but not the transfer matrix.
Only the R-matrix and K-matrix compose the transfer matrix,
and the vertex operators do not appear in.
The roles of them are difference.
Finally, we point out an open problem of the MPA\@.
We have adopted the vertex operator as the ZF-algebra
only because the result must be identical to the known one
in a certain limit.
For application of our MPA,
one needs results of other theories to realize the ZF-algebra.
There are yet no prescription to derive it within a scope of the MPA.

\setcounter{equation}{0}
\section{Realization of Boundary States}
\label{sec:3}

In this section we realize the boundary states
in terms of $U_q(\widehat{sl}_2)$.
The relations (\ref{eqn:def_bo})-(\ref{eqn:def_dual_bo}) request
\begin{eqnarray}
  \mathcal{V}\cdot\Phi(\zeta^{-1})K^+(\zeta)
  =
  \mathcal{V}\cdot\Phi(\zeta),
  &&\qquad
  K^-(\zeta)\Phi(\zeta)\cdot\mathcal{W}
  =
  \Phi(\zeta^{-1})\cdot\mathcal{W}, \\
  K^+(\zeta)\Phi^*(\zeta)\cdot\mathcal{V^*}
  =
  \Phi^*(\zeta^{-1})\cdot\mathcal{V^*},
  &&\qquad
  \mathcal{W^*}\cdot\Phi(\zeta^{-1})K^-(\zeta)
  =
  \mathcal{W^*}\cdot\Phi(\zeta).
\end{eqnarray}
We make the ansatz that they have the forms
\begin{eqnarray}
  \mathcal{V} = \mbox{e}^{F^{(V)}}\otimes \mbox{e}^{-\Lambda-\alpha/2}
  \:\in\: \mathcal{H}^{(1)},
  &&\qquad
  \mathcal{W} = \mbox{e}^{F^{(W)}}\otimes \mbox{e}^\Lambda
  \:\in\: \mathcal{H}^{(0)}, \\
  \mathcal{V}^* = \mbox{e}^{F^{(V^*)}}\otimes \mbox{e}^{\Lambda+\alpha/2}
  \:\in\: \mathcal{H}^{(1)},
  &&\qquad
  \mathcal{W}^* = \mbox{e}^{F^{(W^*)}}\otimes \mbox{e}^{-\Lambda}
  \:\in\: \mathcal{H}^{(0)}.
\end{eqnarray}
The definitions of the spaces $\mathcal{H}^{(i)}$ ($i=0,1$)
and the operators $\mbox{e}^\Lambda$, $\mbox{e}^\alpha$
are given in Appendix \ref{sec:app2}.
Since the spectral parameters are reversed $\zeta\rightarrow\zeta^{-1}$
by the action of the K-matrices,
the prior parts of the boundary states
$F^{(V)}$, $F^{(W)}$, $F^{(V^*)}$ and $F^{(W^*)}$
must have the quadratic terms of bosonic operators.
They are expressed by the following forms,
\begin{eqnarray}
  F^{(V)} &=&
  \frac{1}{2}\sum_{k=1}^{\infty}
  \frac{k\alpha_{k}^{(V)}}{[2k][k]}a_k^2
  + \sum_{k=1}^{\infty}
  \beta_{k}^{(V)}a_k, \\
  F^{(W)} &=&
  \frac{1}{2}\sum_{k=1}^{\infty}
  \frac{k\alpha_{k}^{(W)}}{[2k][k]}a_{-k}^{2}
  + \sum_{k=1}^{\infty}
  \beta_{k}^{(W)}a_{-k}, \\
  F^{(V^*)} &=&
  \frac{1}{2}\sum_{k=1}^{\infty}
  \frac{k\alpha_{k}^{(V^*)}}{[2k][k]}a_{-k}^{2}
  + \sum_{k=1}^{\infty}
  \beta_{k}^{(V^*)}a_{-k}, \\
  F^{(W^*)} &=&
  \frac{1}{2}\sum_{k=1}^{\infty}
  \frac{k\alpha_{k}^{(W^*)}}{[2k][k]}a_k^2
  + \sum_{k=1}^{\infty}
  \beta_{k}^{(W^*)}a_k,
\end{eqnarray}
where $[n]$ is the $q$-integer (see eq. (\ref{eqn:comm_a})).
Noting the following formulas
\begin{eqnarray}
\label{eqn:comm_op1}
  \mbox{e}^A \mbox{e}^B = \mbox{e}^{\frac{1}{2}[A,[A,B]]}\mbox{e}^{[A,B]}\mbox{e}^B \mbox{e}^A,
  \qquad (\mbox{for $[A,[A,B]]\in\textbf{C}$}) \\
\label{eqn:comm_op2}
  \mbox{e}^A \mbox{e}^B = \mbox{e}^B \mbox{e}^A \mbox{e}^{[A,B]}\mbox{e}^{\frac{1}{2}[[A,B],B]},
  \qquad (\mbox{for $[[A,B],B]\in\textbf{C}$})
\end{eqnarray}
and the relation $a_k\cdot 1=1\cdot a_{-k}=0$ ($k>0$),
$\partial\cdot 1=1\cdot\partial=0$,
we find that
\begin{eqnarray}
\label{eqn:bo_a,b_1}
  \alpha_{k}^{(V)} = -\alpha_k,
  &&\qquad
  \beta_{k}^{(V)} = \beta_{k+}, \\
  \alpha_{k}^{(W)} = -\alpha_{k}^{-1},
  &&\qquad
  \beta_{k}^{(W)} = -\alpha_k^{-1}\beta_{k-}, \\
  \alpha_{k}^{(V^*)} = -\alpha_k^*,
  &&\qquad
  \beta_{k}^{(V^*)} = \beta_{k+}^*, \\
\label{eqn:bo_a,b_2}
  \alpha_{k}^{(W^*)} = -\alpha_{k}^{*-1},
  &&\qquad
  \beta_{k}^{(W^*)} = -\alpha_{k}^{*-1}\beta_{k-}^*,
\end{eqnarray}
where
\begin{eqnarray}
  \alpha_k = q^{-6k},
  &&\qquad
  \beta_{k\pm} = \frac{q^{-7k/2}}{[2k]}(\theta_k(1-q^k)+q^k r_\pm^k), \\
  \alpha_k^* = q^{2k},
  &&\qquad
  \beta_{k\pm}^* = -\frac{q^{k/2}}{[2k]}(\theta_k(1-q^k)-q^{-k}r\pm^k),
\end{eqnarray}
and $\theta_k=(1+(-1)^k)/2$.
The relations (\ref{eqn:bo_a,b_1})-(\ref{eqn:bo_a,b_2})
are easily proved by the following identity
(see eq. (\ref{eqn:kappa,inf}) for the notation $(z;p)_\infty$),
\begin{equation}
  \exp\left(-l\sum_{k=1}^{\infty}\frac{[mk]}{k[nk]}z^k\right) =
  \left(\frac{(q^{n-m}z;q^{2n})_\infty}
    {(q^{n+m}z;q^{2n})_\infty}\right)^l.
\end{equation}
The boundary states $\mathcal{W}$ and $\mathcal{W}^*$
are equal to the eigenstates in ref \cite{Jimbo}.
The MPA has been approved.

\setcounter{equation}{0}
\section{Correlation Functions}
\label{sec:4}

To find the correlation functions we compute the following $2N$-point
function,
\begin{eqnarray}
  &&
  P_{\epsilon_1^*,\ldots,\epsilon_N^*;\epsilon_N,\ldots,\epsilon_1}
  (\{\zeta_j^*\};\{\zeta_i\})
  \nonumber \\
  && =
  G\times
  \langle\mathcal{W^*}\cdot
  \Phi_{\epsilon_1^*}^{*(0,1)}(\zeta_1^*)\cdots
  \Phi_{\epsilon_N^*}^{*(1-\iota,\iota)}(\zeta_N^*)
  \cdot\mathcal{V^*}|\mathcal{V}\cdot
  \Phi_{\epsilon_N}^{(\iota,1-\iota)}(\zeta_N)\cdots
  \Phi_{\epsilon_1}^{(1,0)}(\zeta_1)
  \cdot\mathcal{W}\rangle.
\end{eqnarray}
Here $\iota=N\bmod{2}$ and
$G=g^N/\langle\mathcal{W}^*\cdot\mbox{id}\cdot\mathcal{V}^*|
\mathcal{V}\cdot\mbox{id}\cdot\mathcal{W}\rangle$.
Note that the product of the boundary states $\mathcal{V}^*$ and $\mathcal{V}$
can be regarded as an element of
either $\mathcal{H}^{(0)}$ or $\mathcal{H}^{(1)}$
by the cancellation of the operators $\mbox{e}^\alpha$.
First we compute the normal-ordering of the bosonic operators.
By the bosonic formulas (\ref{eqn:comm_a}), (\ref{eqn:comm_der}) and
the expressions of the vertex operators (\ref{eqn:def_ver_minus}) and
(\ref{eqn:def_ver_plus}), we have the following expressions.
\begin{eqnarray}
\label{eqn:phys}
  \Phi_{\epsilon_N}^{(\iota,1-\iota)}(\zeta_N)\cdots
  \Phi_{\epsilon_1}^{(1,0)}(\zeta_1) &=&
  \prod_{a\in A}\oint_{C_a}\frac{\mbox{d}w_a}{2\pi\mbox{i}}
  C(\{\zeta_i\},\{w_a\})\times I(\{\zeta_i\},\{w_a\}), \\
\label{eqn:dual_phys}
  \Phi_{\epsilon_1^*}^{*(0,1)}(\zeta_1^*)\cdots
  \Phi_{\epsilon_N^*}^{*(1-\iota,\iota)}(\zeta_N^*) &=&
  \prod_{b\in B}\oint_{C_b^*}\frac{\mbox{d}w_b^*}{2\pi\mbox{i}}
  C^*(\{\zeta_j^*\},\{w_b^*\})\times I^*(\{\zeta_j^*\},\{w_b^*\}),
\end{eqnarray}
where
\begin{eqnarray}
  C(\{\zeta_i\},\{w_a\}) &=&
  (-q^3)^{\frac{1}{2}\left[\frac{N}{2}\right]
    +\frac{N(N-1)}{4}-\sum_{a\in A}(N-a+1)}
  \prod_{i=1}^{N}\zeta_{i}^{i-1+\frac{1+\epsilon_i}{2}}
  \prod_{i>i'}\frac{(q^2\zeta_{i'}^2/\zeta_i^2;q^4)_\infty}
  {(q^4\zeta_{i'}^2/\zeta_i^2;q^4)_\infty}
  \nonumber \\
  && \qquad \times
  \frac{\prod_{a\in A}q^{-1}(1-q^2)w_a\prod_{a>a'}(w_a-w_{a'})(w_a-q^2 w_{a'})}
  {\prod_{i\geq a}(\zeta_i^2-q^{-2}w_a)
    \prod_{i\leq a}(w_a-q^4\zeta_i^2)}, \\
  \nonumber \\
  C^*(\{\zeta_j^*\},\{w_b^*\}) &=&
  (-q^3)^{\frac{1}{2}\left[\frac{N}{2}\right]
    +\frac{N(N-1)}{4}-\sum_{b\in B}b}
  \prod_{j=1}^{N}(-q^{-1}\zeta_j^*)^{N-j+\frac{1-\epsilon_j^*}{2}}
  \prod_{j<j'}\frac{(q^2\zeta_{j'}^{*\: 2}/\zeta_{j}^{*\: 2};q^4)_\infty}
  {(q^4\zeta_{j'}^{*\: 2}/\zeta_{j}^{*\: 2};q^4)_\infty}
  \nonumber \\
  && \qquad \times
  \frac{\prod_{b\in B}q^{-1}(1-q^2)w_b^*
    \prod_{b<b'}(w_b^*-w_{b'}^*)(w_b^*-q^2 w_{b'}^*)}
  {\prod_{j\leq b}q^{-2}(\zeta_j^{*\: 2}-w_b^*)
    \prod_{j\geq b}(w_b^*-q^2\zeta_j^{*\: 2})}, \\
  \nonumber \\
  I(\{\zeta_i\},\{w_a\}) &=&
  \exp\left(\sum_{k=1}^{\infty}a_{-k}X_k(\{\zeta_i\},\{w_a\})\right)
  \exp\left(-\sum_{k=1}^{\infty}a_k Y_k(\{\zeta_i\},\{w_a\})\right)
  \nonumber \\
  && \qquad \otimes
  \mbox{e}^{-\sum_{i=1}^N \epsilon_i \alpha/2}  
  \prod_{i=1}^{N}(-q^3\zeta_i^2)^{\partial/2}
  \prod_{a\in A}w_{a}^{-\partial}, \\
  I^*(\{\zeta_j^*\},\{w_b^*\}) &=&
  \exp\left(\sum_{k=1}^{\infty}a_{-k}X^*_k(\{\zeta_j^*\},\{w_b^*\})\right)
  \exp\left(-\sum_{k=1}^{\infty}a_k Y_k^*(\{\zeta_j^*\},\{w_b^*\})\right)
  \nonumber \\
  && \qquad\otimes
  \mbox{e}^{\sum_{j=1}^N \epsilon_j^* \alpha/2}
  \prod_{i=1}^{N}(-q\zeta_{j}^{*\: 2})^{\partial/2}
  \prod_{b\in B}w_{b}^{*-\partial},
\end{eqnarray}
with
\begin{eqnarray}
  X_k(\{\zeta_i\},\{w_a\}) &=&
  \frac{q^{7k/2}}{[2k]}\sum_{i=1}^{N}\zeta_{i}^{2k}
  -\frac{q^{k/2}}{[k]}\sum_{a\in A}w_a^k, \\
  Y_k(\{\zeta_i\},\{w_a\}) &=&
  \frac{q^{-5k/2}}{[2k]}\sum_{i=1}^{N}\zeta_{i}^{-2k}
  -\frac{q^{k/2}}{[k]}\sum_{a\in A}w_{a}^{-k}, \\
  X_k^*(\{\zeta_j^*\},\{w_b^*\}) &=&
  \frac{q^{3k/2}}{[2k]}\sum_{j=1}^{N}\zeta_{j}^{*\: 2k}
  -\frac{q^{k/2}}{[k]}\sum_{b\in B}w_{b}^{*\: k}, \\
  Y_k^*(\{\zeta_j^*\},\{w_b^*\}) &=&
  \frac{q^{-k/2}}{[2k]}\sum_{j=1}^{N}\zeta_{j}^{*-2k}
  -\frac{q^{k/2}}{[k]}\sum_{b\in B}w_{b}^{*-k}.
\end{eqnarray}
In the above, sets $A$ and $B$ are defined by
\begin{equation}
  A = \{i|\epsilon_i=+\},
  \qquad
  B = \{j|\epsilon_j^*=-\}.
\end{equation}
The integration contours $C_a$ and $C_b^*$
in eqs. (\ref{eqn:phys}) and (\ref{eqn:dual_phys})
are chosen to encircle the only points
$q^4\zeta_i$ ($i\leq a$) and $q^2\zeta_j^*$ ($j\geq b$) respectively.
By the formulas (\ref{eqn:comm_op1}) and (\ref{eqn:comm_op2}),
it follows that
\begin{eqnarray}
  &&
  |\mathcal{V}\cdot I(\{\zeta_i\},\{w_a\})\cdot\mathcal{W}\rangle
  \nonumber \\
  && \quad =\:
  \exp\left\{
    -\sum_{k=1}^{\infty}\frac{[2k][k]}{k}
    \left(\frac{1}{2\alpha_k}Y_k^2-\frac{\beta_{k-}}{\alpha_k}Y_k
    \right)\right\}
  \nonumber \\
  &&\qquad\times
  |\mbox{e}^{F^{(V)}}\mbox{e}^{F^{(W)}}
  \exp\left(\sum_{k=1}^{\infty}a_{-k}(X_k+\alpha_{k}^{-1}Y_k)\right)
  \exp\left(-\sum_{k=1}^{\infty}a_k Y_k\right)
  \nonumber \\
  &&\qquad\quad\otimes
  \mbox{e}^{-\frac{\alpha}{2}(\sum_{i=1}^N\epsilon_i+1)}
  \prod_{i=1}^{N}(-q^3 \zeta_i^2)^{\partial/2}
  \prod_{a\in A}w_{a}^{-\partial}\rangle, \\
  &&
  \langle \mathcal{W}^*\cdot I^*(\{\zeta_j^*\},\{w_b^*\})\cdot\mathcal{V}^*|
  \nonumber \\
  && \quad =\:
  \exp\left\{
    -\sum_{k=1}^{\infty}\frac{[2k][k]}{k}
    \left(\frac{\alpha_k^*}{2}Y_k^{*\: 2}
      +\beta_{k+}^*Y_k^*\right)\right\}
  \nonumber \\
  &&\qquad\quad\times
  \prod_{j=1}^{N}(-q\zeta_{j}^{*\: 2})^{1/2}
  \prod_{b\in B}w_{b}^{*-1}
  \nonumber \\
  &&\qquad\times
  \langle \mbox{e}^{F^{(W^*)}} \mbox{e}^{F^{(V^*)}}
  \exp\left(\sum_{k=1}^{\infty}a_{-k}(X_k^*+\alpha_k^* Y_k^*)\right)
  \exp\left(-\sum_{k=1}^{\infty}a_k Y_k^*\right)
  \nonumber \\
  &&\qquad\quad\otimes
  \mbox{e}^{\frac{\alpha}{2}(\sum_{j=1}^N\epsilon_j^*+1)}
  \prod_{i=1}^{N}(-q\zeta_{j}^{*\: 2})^{\partial/2}
  \prod_{b\in B}w_{b}^{*-\partial}|.
\end{eqnarray}
Combining these expressions together we obtain
\begin{eqnarray}
\label{eqn:comb}
  &&
  P_{\epsilon_1^*,\ldots,\epsilon_N^*;\epsilon_N,\ldots,\epsilon_1}
  (\{\zeta_j^*\};\{\zeta_i\})
  \nonumber \\
  && \quad =\:
  \delta_{\epsilon\epsilon^*}
  \prod_{a\in A}\prod_{b\in B}\oint_{C_a}\oint_{C_b^*}
  \frac{\mbox{d}w_a}{2\pi\mbox{i}}\frac{\mbox{d}w_b^*}{2\pi\mbox{i}}
  C(\{\zeta_i\},\{w_a\})C^*(\{\zeta_j^*\},\{w_b^*\})
  \nonumber \\
  && \qquad\times
  \left(\prod_{j=1}^{N}(-q\zeta_{j}^{*\: 2})
    \prod_{b\in B}w_{b}^{*-2}\right)^{-\frac{1}{2}(\sum_{i=1}^N\epsilon_i+1)}
  \hspace{-1cm}\times
  J(\{\zeta_i\},\{w_a\};\{\zeta_j^*\},\{w_b^*\}),
\end{eqnarray}
where
\begin{eqnarray}
\label{eqn:def_J}
  &&
  J(\{\zeta_i\},\{w_a\};\{\zeta_j^*\},\{w_b^*\})
  \nonumber \\
  && \quad=\:
  G\times\exp\left\{-\sum_{k=1}^{\infty}\frac{[2k][k]}{k}
    \left(\frac{1}{2\alpha_k}Y_k^2+\frac{\alpha_k^*}{2}Y_k^{*\: 2}
      -\frac{\beta_{k-}}{\alpha_k}Y_k+\beta_{k+}^*Y_k^*\right)\right\}
  \nonumber \\
  && \qquad\times
  \langle \mbox{e}^{F^{(W^*)}}\mbox{e}^{F^{(V^*)}}
  \exp(\mbox{$\sum_{k=1}^{\infty}$}a_{-k}(X_k^*+\alpha_k^* Y_k^*)
  \exp(-\mbox{$\sum_{k=1}^{\infty}$}a_k Y_k^*))|
  \nonumber \\
  && \quad\qquad\times
  |\mbox{e}^{F^{(V)}}\mbox{e}^{F^{(W)}}\exp(\mbox{$\sum_{k=1}^{\infty}$}
  a_{-k}(X_k+\alpha_{k}^{-1}Y_k))\rangle.
\end{eqnarray}
The Kronecker's delta $\delta_{\epsilon\epsilon^*}$
in eq. (\ref{eqn:comb}) takes a value $1$
in the case $\sum\epsilon_i=\sum\epsilon_i^*$ and vanishes otherwise.
Next, we compute $J(\{\zeta_i\},\{w_a\};\{\zeta_j^*\},\{w_b^*\})$.
We insert the completeness relation
\begin{eqnarray}
  &&
  \mbox{id} =
  \int\prod_{k=1}^{\infty}\frac{k\mbox{d}\xi_k\mbox{d}\bar{\xi}_k}{[2k][k]}
  \exp\left(-\sum_{k=1}^{\infty}\frac{k}{[2k][k]}\xi_k\bar{\xi}_k\right)
  \exp\left(\sum_{k=1}^{\infty}\frac{k}{[2k][k]}\xi_k a_{-k}\right)
  \nonumber \\
  && \qquad\qquad\times
  |1\otimes \mbox{e}^\Lambda\rangle\langle1\otimes \mbox{e}^{-\Lambda}|
  \exp\left(\sum_{k=1}^{\infty}\frac{k}{[2k][k]}\bar{\xi}_k a_k\right),
\end{eqnarray}
between $\mbox{e}^{F^{(W^*)}}$ and $\mbox{e}^{F^{(V^*)}}$,
and between $\mbox{e}^{F^{(V)}}$ and $\mbox{e}^{F^{(W)}}$
in the last bracket of eq. (\ref{eqn:def_J}).
Then we have the following integral expression,
\begin{eqnarray}
\label{eqn:J_cal}
  &&
  J(\{\zeta_i\},\{w_a\};\{\zeta_j^*\},\{w_b^*\}) =
  G\times\int\prod_{k=1}^{\infty}\frac{k}{[2k][k]}
  \mbox{d}\xi_k\mbox{d}\bar{\xi}_k\mbox{d}\xi_k^*\mbox{d}\bar{\xi}_k^*
  \nonumber \\
  && \quad\times
  \exp\biggl\{-\frac{1}{2}\sum_{k=1}^{\infty}\frac{k}{[2k][k]}
    [\xi_k,\bar{\xi}_k,\xi_k^*,\bar{\xi}_k^*]\mathcal{A}_k
    \left[
      \begin{array}{c}
        \xi_k \\
        \bar{\xi}_k \\
        \xi_k^* \\
        \bar{\xi}_k^*
      \end{array}
    \right]
    +\sum_{k=1}^{\infty}[\xi_k,\bar{\xi}_k,\xi_k^*,\bar{\xi}_k^*]
    \mathcal{B}_k\biggr\}
  \nonumber \\
  && \quad\times
  \exp\left\{-\sum_{k=1}^{\infty}\frac{[2k][k]}{k}
    \left(\frac{1}{2\alpha_k}Y_k^2+\frac{\alpha_k^*}{2}Y_k^{*\: 2}
      -\frac{\beta_{k-}}{\alpha_k}Y_k+\beta_{k+}^*Y_k^*\right)\right\},
\end{eqnarray}
where the matrix $\mathcal{A}_k$ and the vector $\mathcal{B}_k$ are given by
\begin{equation}
  \mathcal{A}_k =
  \left[
    \begin{array}{cccc}
      \alpha_k & 1 & 0 & -1 \\
      1 & \alpha_{k}^{-1} & 0 & 0 \\
      0 & 0 & \alpha_{k}^{*-1} & 1 \\
      -1 & 0 & 1 & \alpha_k^*
    \end{array}
  \right],
  \qquad
  \mathcal{B}_k =
  \left[
    \begin{array}{l}
      \beta_{k+}-Y_k^* \\
      -\alpha_{k}^{-1}\beta_{k-}+X_k+\alpha_{k}^{-1}Y_k \\
      -\alpha_{k}^{*-1}\beta_{k-}^{*} \\
      \beta_{k+}^{*}+X_k^*+\alpha_k^* Y_k^*
    \end{array}
  \right].
\end{equation}
Using the formula for the Gaussian integral
\begin{eqnarray}
  &&
  \int\prod_{k=1}^{\infty}\frac{k}{[2k][k]}
  \mbox{d}\xi_k\mbox{d}\bar{\xi}_k\mbox{d}\xi_k^*\mbox{d}\bar{\xi}_k^*
  \nonumber \\
  && \quad\times
  \exp\biggl\{-\frac{1}{2}\sum_{k=1}^{\infty}\frac{k}{[2k][k]}
    [\xi_k,\bar{\xi}_k,\xi_k^*,\bar{\xi}_k^*]\mathcal{A}_k
    \left[
      \begin{array}{c}
        \xi_k \\
        \bar{\xi}_k \\
        \xi_k^* \\
        \bar{\xi}_k^*
      \end{array}
    \right]
    +\sum_{k=1}^{\infty}[\xi_k,\bar{\xi}_k,\xi_k^*,\bar{\xi}_k^*]
    \mathcal{B}_k\biggr\}
    \nonumber \\
    && =
    \prod_{k=1}^{\infty}(-\det\mathcal{A}_k)^{-1/2}
    \exp\left(\frac{1}{2}\sum_{k=1}^{\infty}\frac{[2k][k]}{k}
      \mathcal{B}_k^t\mathcal{A}_k\mathcal{B}_k\right),
\end{eqnarray}
in eq. (\ref{eqn:J_cal}), we get
\begin{eqnarray}
\label{eqn:J_ab}
  &&
  J(\{\zeta_i\},\{w_a\};\{\zeta_j^*\},\{w_b^*\})
  \nonumber \\
  && \quad = \:
  g^N\exp\left\{\sum_{k=1}^{\infty}\frac{[2k][k]}{k}
    \biggl(\frac{1}{2}\alpha_k X_k^2+\frac{1}{2}\alpha_k^*Y_{k}^{*\: 2}
  \right. \nonumber \\
  && \qquad
  +(\alpha_k(\beta_{k+}^{*}+\beta_{k-}^{*})-\beta_{k-})X_k
  -(\beta_{k+}+\beta_{k-})X_k^*
  +(\beta_{k+}^{*}+\beta_{k-}^{*})Y_k
  \nonumber \\ 
  && \qquad
  -(\alpha_k^*(\beta_{k+}+\beta_{k-})-\beta_{k-}^{*})Y_k^*
  +\alpha_k X_k(X_k^*+\alpha_k^*Y_k^*)
  \nonumber \\
  && \quad\qquad\left.
  +Y_k(X_k+X_k^*)
  +Y_k^*(X_k^*+\alpha_k^*Y_k)
  \biggr)\right\}.
\end{eqnarray}
We have used the definition of the normalization factor $G$
($=g^N/\langle\mathcal{W}^*\cdot\mbox{id}\cdot\mathcal{V}^*|
\mathcal{V}\cdot\mbox{id}\cdot\mathcal{W}\rangle$).
From eq.~(\ref{eqn:comb}) with eq.~(\ref{eqn:J_ab})
we arrive at
\begin{eqnarray}
\label{eqn:result}
  &&
  P_{\epsilon_1^*,\ldots,\epsilon_N^*;\epsilon_N,\ldots,\epsilon_1}
  (\{\zeta_j^*\};\{\zeta_i\})
  \nonumber \\
  && =
  \delta_{\epsilon\epsilon^*}g^N
  \prod_{a\in A}\prod_{b\in B}\oint_{C_a}\oint_{C_b^*}
  \frac{\mbox{d}w_a}{2\pi\mbox{i}}\frac{\mbox{d}w_b^*}{2\pi\mbox{i}}
  C(\{\zeta_i\},\{w_a\})C^*(\{\zeta_j^*\},\{w_b^*\})
  \nonumber \\
  &&\quad\times
  \left(\prod_{j=1}^{N}(-q\zeta_{j}^{*\: 2})
    \prod_{b\in B}w_{b}^{*-2}\right)^{-\frac{1}{2}(\sum_{i=1}^N \epsilon_i+1)}
  \nonumber \\
  &&\quad\times
  \prod_{i,j=1}^{N}\prod_{a\in A\atop{b\in B}}\left(
  \frac{\phi_{-5/2,+}^{*}(\zeta_i^2)\psi_{-11/2,+}^{*}(w_a)
    \phi_{-5/2,-}^{*}(\zeta_i^2)\psi_{-11/2,-}^{*}(w_a)}
  {\phi_{7/2,-}(\zeta_i^2)\psi_{1/2,-}(w_a)}\right.
  \nonumber \\
  &&\quad\qquad\times
  \frac{\phi_{-5/2,+}^{*}(\zeta_{i}^{-2})\psi_{1/2,+}^{*}(w_{a}^{-1})
    \phi_{-5/2,-}^{*}(\zeta_{i}^{-2})\psi_{1/2,-}^{*}(w_{a}^{-1})}
  {\phi_{3/2,+}(\zeta_{j}^{*\: 2})\psi_{1/2,+}(w_b^*)
    \phi_{3/2,-}(\zeta_{j}^{*\: 2})\psi_{1/2,-}(w_b^*)}
  \nonumber \\
  &&\quad\qquad\times\left.
    \frac{\phi_{-1/2,-}^{*}(\zeta_{j}^{*-2})\psi_{1/2,-}^{*}(w_{b}^{*-1})}
    {\phi_{3/2,+}(\zeta_{j}^{*-2})\psi_{5/2,+}(w_{b}^{*-1})
      \phi_{3/2,-}(\zeta_{j}^{*-2})\psi_{5/2,-}(w_{b}^{*-1})}
  \right)
  \nonumber \\
  &&\quad\times
  \biggl(\tau_{XX,-6}(\{\zeta_i\},\{w_a\})
  \tau_{Y^* Y^*,2}(\{\zeta_j^*\},\{w_b^*\})\biggr)^{1/2}
  \nonumber \\
  &&\qquad\times
  \tau_{XX^*,-6}(\{\zeta_i\},\{w_a\};\{\zeta_j^*\},\{w_b^*\})
  \tau_{XY^*,-4}(\{\zeta_i\},\{w_a\};\{\zeta_j^*\},\{w_b^*\})
  \nonumber \\
  &&\qquad\times
  \tau_{YY^*,2}(\{\zeta_i\},\{w_a\};\{\zeta_j^*\},\{w_b^*\})
  \tau_{YX^*,0}(\{\zeta_i\},\{w_a\};\{\zeta_j^*\},\{w_b^*\})
  \nonumber \\
  &&\qquad\times
  \tau_{XY,0}(\{\zeta_i\},\{w_a\})
  \tau_{X^* Y^*,0}(\{\zeta_j^*\},\{w_b^*\}).
\end{eqnarray}
The functions $\phi_{a\pm}(\zeta)$, $\psi_{a\pm}(w)$, $\tau(\{\zeta\},\{w\})$
are given in Appendix \ref{sec:app3}. 
The integration contours $C_a$ and $C_b^*$ encircle the following points,
\begin{eqnarray}
  C_a &:&
  q^4\zeta_i^2 ; r_\pm^{-1}, q^6 r_\pm^{-1} ; 0, \pm q^2, \pm q^4 ;
  w_{a'}, q^{-2}w_{a'} ;
  \nonumber \\
  &&
  q^6 w_b^{*-1}, q^4 w_b^{*-1}, q^4 w_b^*, w_b^*, q^{-2}w_b^*, q^{-4}w_b^* ;
  \nonumber \\
  &&  
  \mbox{($b\in B$, $a'\in A/\{a\}$ and $i\leq a$)},
  \\
  C_b^* &:&
  q^2\zeta_j^{*\: 2} ; r_-, q^2 r_\pm^{-1} ; 0, \pm q^{-1}, \pm q^3 ;
  w_{b'}^*, q^2 w_{b'}^* ;
  \nonumber \\
  &&
  q^6 w_a^{-1}, q^4 w_a^{-1} ; q^4 w_a, q^2 w_a, w_a, q^{-4}w_a;
  \nonumber \\
  &&  
  \mbox{($a\in A$, $b'\in B/\{a\}$ and $j\geq b$)}.
\end{eqnarray}
See eqs. (\ref{eqn:form1}) and (\ref{eqn:form2})
for the definitions of $\tau_{XY,\alpha}$ and $\tau_{X^* Y^*,\alpha}$.
While $\tau_{XY,0}$ and $\tau_{X^* Y^*,0}$ diverge at $a=a'$ and $b=b'$,
the residues at the points $q^4 \zeta_i$ and $q^2 \zeta_j^*$ are respectively
finite.
The other functions are always finite.

The correlation function including the local operator $\mathcal{O}$
is given by
\begin{eqnarray}
\label{eqn:result_corr}
  &&
  \langle\mathcal{O}\rangle =
  \sum_{m=1}^{N}\sum_{\epsilon_m,\epsilon'_m=\pm}
  (\mathcal{O}_m)_{\epsilon_m \epsilon'_m}
  E_{m}^{\epsilon_m \epsilon'_m}, \\
  &&
  E_{m}^{\epsilon_m\epsilon'_m} =
  \sum_{\mbox{\scriptsize except for }\epsilon_m,\epsilon'_m}
  P_{\epsilon_1,\ldots,\epsilon_m,\ldots,\epsilon_N;
    \epsilon_N,\ldots,\epsilon'_m,\ldots,\epsilon_1}
  (\{1\};\{1\}).
\end{eqnarray}
For instance, the magnetization on a site $m$ is
\begin{equation}
  \langle\sigma_m^z\rangle =  E_{m}^{++} - E_{m}^{--}.
\end{equation}

\setcounter{equation}{0}
\section{Discussion}
\label{sec:5}

We have applied the MPA to the finite XXZ model with boundaries
and calculated the $2N$-point function.
The $2N$-point function (\ref{eqn:result}) and
the correlation function (\ref{eqn:result_corr}) are the main results
of this paper.
The type I vertex operator is regarded as the ZF-algebra
and the boundary states are realized
in the bosonic $U_q(\widehat{sl}_2)$ form. 
The eigenstate is given by the product of them.
We have set the vertex operators free from those physical meanings,
observed those mathematical properties,
and succeeded in analyzing the model with boundaries.
Such manipulations are allowed only in the MPA\@.
The vertex operators constitute the eigenstate and do not participate
in the transfer matrix.
The transfer matrix is made of the R-matrix and the boundary K-matrix.

There are open problems.
It is interesting to look for a closed equation
for the correlation function.
For the infinite XXZ model
the correlation function is known to satisfy a certain difference equation
called the quantum Knizhnik-Zamolodchikov
equation \cite{CBMS,Frenkel,Cherednik_KZ,Matsuo,q-KZ},
and the similar equation is found also for the half-infinite case \cite{Miwa}.
In the finite chain case similar properties are expected.
Solving that equation may give a practical method
to calculate observable quantities.
In this viewpoint it is important to investigate further
the form factors \cite{Smirnov}.

It is also interesting to consider how to construct the excited states.
In the infinite case the type II vertex operator plays its role.
However, in the MPA, the situation is not so simple.
The excited states are known to be obtained by the product of the ZF-algebra,
the boundary states, and other operators
that satisfy certain relations \cite{Sasamoto}.
We have to find the third operators
and show the relations among them and the type II vertex operators.

It is possible and important to apply the MPA to other models,
for instance, the massless XXZ model.
Recently, the elliptic quantum algebra has been developed \cite{Ellip,Iohara}
and the vertex operators for the degenerate elliptic algebra
have been realized in the bosonic form \cite{Konno}.
They may help us to analyze the massless XXZ model with boundaries.

Finally, we point out a connection between the MPA and the ``affinization''.
In our MPA, the realization of the ZF-algebra has been justified by comparing
with the known result.
We like to define the ZF-algebra uniquely within a scope of the MPA\@.
Note that the definition of the ZF-algebra physically
means the Yang-Baxter equation for the transfer matrix
on the half-infinite lattice.
Such a transfer matrix is obtained by extending the original one
into the infinite chain limit, i.e.\ by the ``affinization''
of symmetry of the model.
We expect that the ZF-algebra is uniquely realized by the affinization
of the transfer matrix.
The MPA may be mathematically defined by using the idea of the affinization.

\section*{Acknowledgment}
The authors would like to thank T. Sasamoto, M. Shiroishi and
T. Takagi for fruitful discussions and comments.

\appendix
\renewcommand{\theequation}{\Alph{section}.\arabic{equation}}
\section{R-Matrix and Boundary K-Matrix}
\label{sec:app1}

We summarize some basic formulas for the R-matrix and the boundary K-matrix
of the XXZ model.

\subsection{R-Matrix}
\label{sec:app1.1}

The R-matrix is
\begin{equation}
  R(\zeta) =
  \frac{1}{\kappa(\zeta)}
  \left[
    \begin{array}{cccc}
      1 &&& \\
      & \frac{\displaystyle (1-\zeta^2)q}{\displaystyle 1-q^2\zeta^2} &
      \frac{\displaystyle (1-q^2)\zeta}{\displaystyle 1-q^2\zeta^2} & \\
      & \frac{\displaystyle (1-q^2)\zeta}{\displaystyle 1-q^2\zeta^2} &
      \frac{\displaystyle (1-\zeta^2)q}{\displaystyle 1-q^2\zeta^2} & \\
      &&& 1
    \end{array}
  \right],
\end{equation}
where
\begin{equation}
\label{eqn:kappa,inf}
  \kappa(\zeta) =
  \frac{(q^4\zeta^2;q^4)_\infty(q^2\zeta^{-2};q^4)_\infty}
  {(q^4\zeta^{-2};q^4)_\infty(q^2\zeta^2;q^4)_\infty}\zeta,
  \qquad
  (z;p)_\infty = \prod_{n=0}^{\infty}(1-zp^n).
\end{equation}
The R-matrix satisfies the Yang-Baxter equation, the unitarity
and the crossing symmetry,
\begin{eqnarray}
\label{eqn:YBE}
  R_{12}(\zeta_1/\zeta_2)R_{13}(\zeta_1/\zeta_3)R_{23}(\zeta_2/\zeta_3)
  &=&
  R_{23}(\zeta_2/\zeta_3)R_{13}(\zeta_1/\zeta_3)R_{12}(\zeta_1/\zeta_2), \\
  R_{12}(\zeta_1/\zeta_2)R_{21}(\zeta_2/\zeta_1) &=& 1, \\
  R_{\epsilon_1\epsilon_2}^{\epsilon'_1\epsilon'_2}(\zeta_1/\zeta_2)
  &=&
  R_{-\epsilon'_2\epsilon_1}^{-\epsilon_2\epsilon'_1}(-q^{-1}\zeta_2/\zeta_1).
\end{eqnarray}

\subsection{Boundary K-Matrix}
\label{sec:app1.2}

In this paper, we choose the diagonal K-matrix,
\begin{equation}
  K(\zeta;r) =
  \frac{1}{f(\zeta;r)}
  \left[
    \begin{array}{cc}
      \frac{\displaystyle 1-r\zeta^2}{\displaystyle \zeta^2-r} & \\
      & 1 
    \end{array}
  \right],
\end{equation}
where
\begin{equation}
  f(\zeta;r) =
  \frac{\varphi(\zeta^{-2};r)}{\varphi(\zeta^2;r)},
  \qquad
  \varphi(z;r) =
  \frac{(q^4 rz;q^4)_\infty(q^6 z^2;q^8)_\infty}
  {(q^2 rz;q^4)_\infty(q^8 z^2;q^8)_\infty}.
\end{equation}
In the text, we denote $K(\zeta;r_\pm)$ by $K^\pm(\zeta)$.
The K-matrix has the following properties,
\begin{eqnarray}
  K(\zeta;r)K(\zeta^{-1};r) &=& 1, \\
  K_a^b(-q^{-1}\zeta^{-1};r) &=&
  \sum_{a',b'}R_{a' -b'}^{-a\: b}(-q\zeta^2)K_{b'}^{a'}(\zeta;r).
\end{eqnarray}

\setcounter{equation}{0}
\section{Vertex Operators}
\label{sec:app2}

We formulate the type I vertex operators.
(The type II vertex operators are omitted since they are not used
in this paper.)
We consider the bosonic Fock space
\begin{equation}
  \mathcal{H}^{(i)} =
  \mbox{\textbf{C}}[a_{-1},\ldots]\otimes
  (\oplus_{n\in\mbox{\scriptsize\bf Z}}\mbox{\textbf{C}}
  \mbox{e}^{\Lambda+n\alpha+\delta_{i1}\alpha/2}),
\end{equation}
for $i=0,1$.
The commutation relations among the bosonic operators $\{a_n\}$
($n\in$\textbf{Z}$/\{0\}$) are
\begin{equation}
\label{eqn:comm_a}
  [a_m,a_n] =
  \delta_{m+n,0}\frac{[2m][m]}{m},
  \qquad
  [n] =
  \frac{q^n-q^{-n}}{q-q^{-1}}.
\end{equation}
The operators $\mbox{e}^\alpha$ and $z^\partial$ act as
\begin{equation}
\label{eqn:comm_der}
  \mbox{e}^\alpha\cdot \mbox{e}^\beta = \mbox{e}^{\alpha+\beta},
  \qquad
  z^\partial\cdot \mbox{e}^\alpha = z^{[\partial,\alpha]}\mbox{e}^\alpha\cdot z^\partial,
\end{equation}
where $[\partial,\alpha]=2$ and $[\partial,\Lambda]=0$.
These operators act also on the dual Fock space $\mathcal{H}^{*(i)}$ as
\begin{equation}
  \mbox{e}^\beta\cdot \mbox{e}^\alpha = \mbox{e}^{\beta+\alpha},
  \qquad
  \mbox{e}^\alpha\cdot z^\partial = z^\partial\cdot \mbox{e}^\alpha z^{[\alpha,\partial]}.
\end{equation}
We further define that $\langle \mbox{e}^\alpha\rangle=0$ and
$\langle z^\partial\rangle=1$.

The vertex operators are defined as the intertwiners
that have the following map,
\begin{eqnarray}
\label{eqn:def_ver_map}
  &&
  \Phi^{(1-i,i)}(\zeta):
  \mathcal{H}^{(i)}\rightarrow\mathcal{H}^{(1-i)}\otimes V,
  \nonumber \\
  &&
  \Phi^{(1-i,i)}(\zeta)=
  \sum_{\epsilon=\pm}\Phi_{\epsilon}^{(1-i,i)}(\zeta)\otimes v_\epsilon,
\end{eqnarray}
and are realized in the following bosonic form,
\begin{eqnarray}
\label{eqn:def_ver_minus}
  \Phi_{-}^{(1-i,i)}(\zeta)
  &=&
  \mbox{e}^{P(\zeta)}\mbox{e}^{Q(\zeta)}\otimes
  \mbox{e}^{\alpha/2}(-q^3\zeta^2)^{(\partial+i)/2}\zeta^{-i}, \\
\label{eqn:def_ver_plus}
  \Phi_{+}^{(1-i,i)}(\zeta)
  &=&
  \oint_{C}\frac{\mbox{d}w}{2\pi\mbox{i}}
  \frac{(1-q^2)w\zeta}{q(w-q^2\zeta^2)(w-q^4\zeta^2)}
  \nonumber \\
  &\times&
  \mbox{e}^{P(\zeta)+R(w)}\mbox{e}^{Q(\zeta)+S(w)}\otimes
  \mbox{e}^{-\alpha/2}(-q^3\zeta^2)^{(\partial+i)/2}w^{-\partial}\zeta^{-i},
\end{eqnarray}
where
\begin{eqnarray}
  &&
  P(\zeta) =
  \sum_{k=1}^{\infty}\frac{a_{-k}}{[2k]}q^{7k/2}\zeta^{2k},
  \qquad
  Q(\zeta) =
  -\sum_{k=1}^{\infty}\frac{a_k}{[2k]}q^{-5k/2}\zeta^{-2k}, \\
  &&
  R(w) =
  -\sum_{k=1}^{\infty}\frac{a_{-k}}{[k]}q^{k/2}w^k,
  \qquad
  S(w) =
  \sum_{k=1}^{\infty}\frac{a_k}{[k]}q^{k/2}w^{-k}.
\end{eqnarray}
The integration contour $C$ in eq. (\ref{eqn:def_ver_plus})
encircles the point $q^4\zeta^2$
but not the point $q^2\zeta^2$.

Similarly the dual vertex operators are defined as
\begin{eqnarray}
\label{eqn:def_d_ver_map}
  &&
  \Phi^{*(1-i,i)}(\zeta):
  \mathcal{H}^{(i)}\otimes V\rightarrow\mathcal{H}^{(1-i)},
  \nonumber \\
  &&
  \Phi^{*(1-i,i)}(\zeta)=
  \sum_{\epsilon=\pm}\Phi_{\epsilon}^{*(1-i,i)}(\zeta)\otimes v_\epsilon^*,
\end{eqnarray}
where $v^*$ is the dual vector of $v$ that satisfies
$v_\epsilon^*\cdot v_{\epsilon'}=\delta_{\epsilon\epsilon'}$.
The dual vertex operator is related with the original one through
\begin{equation}
  \Phi_\epsilon^{*(1-i,i)}(\zeta) =
  \Phi_{-\epsilon}^{(1-i,i)}(-q^{-1}\zeta).
\end{equation}
The label $(1-i,i)$ of the vertex operator is omitted
if there is no fear of confusion.
The following relations are known,
\begin{eqnarray}
  &&
  \sum_{\epsilon_1,\epsilon_2=\pm}
  R_{\epsilon'_1 \epsilon'_2}^{\epsilon_1 \epsilon_2}(\zeta_1/\zeta_2)
  \Phi_{\epsilon_1}(\zeta_1)\Phi_{\epsilon_2}(\zeta_2)
  =
  \Phi_{\epsilon'_2}(\zeta_2)\Phi_{\epsilon'_1}(\zeta_1), \\
\label{eqn:inv}
  && \quad\qquad
  g\sum_{\epsilon=\pm}\Phi_\epsilon^*(\zeta)\Phi_\epsilon(\zeta) =
  \mbox{id}
\end{eqnarray}
where $g$ is a scalar constant,
\begin{equation}
\label{eqn:g}
  g = \frac{(q^2;q^4)_\infty}{(q^4;q^4)_\infty}.
\end{equation}

\setcounter{equation}{0}
\section{Formulas}
\label{sec:app3}

We summarize the formulas that are used
for the derivation of eq. (\ref{eqn:result}) from eq. (\ref{eqn:J_ab}).
\begin{eqnarray}
  \phi_{\alpha\pm}(z) &=&
  \exp\left(\sum_{k=1}^{\infty}\frac{[k]}{k}q^{\alpha k}z^k\beta_{k\pm}\right)
  \nonumber \\
  &=&
  \left(\frac{(q^{2\alpha-1}z^2;q^8)_\infty(q^{2\alpha-3}z^2;q^8)_\infty}
    {(q^{2\alpha-5}z^2;q^8)_\infty(q^{2\alpha+1}z^2;q^8)_\infty}\right)^{1/2}
  \frac{(q^{\alpha+1/2}r_\pm z;q^4)_\infty}
  {(q^{\alpha-3/2}r_\pm z;q^4)_\infty}, \\
  \psi_{\alpha\pm}(w) &=&
  \exp\left(\sum_{k=1}^{\infty}
    \frac{[2k]}{k}q^{\alpha k}w^k\beta_{k\pm}\right)
  \nonumber \\
  &=&
  \left(\frac{1-q^{2\alpha-7}w^2}{1-q^{2\alpha-5}w^2}\right)^{1/2}
  \frac{1}{1-q^{\alpha-5/2}r_\pm w}, \\
  \phi_{\alpha\pm}^{*}(z) &=&
  \exp\left(\sum_{k=1}^{\infty}
    \frac{[k]}{k}q^{\alpha k}z^k\beta_{k\pm}^{*}\right)
  \nonumber \\
  &=&
  \left(\frac{(q^{2\alpha+3}z^2;q^8)_\infty(q^{2\alpha+9}z^2;q^8)_\infty}
    {(q^{2\alpha+7}z^2;q^8)_\infty(q^{2\alpha+5}z^2;q^8)_\infty}\right)^{1/2}
  \frac{(q^{\alpha+5/2}r_\pm z;q^4)_\infty}
  {(q^{\alpha+1/2}r_\pm z;q^4)_\infty}, \\
  \psi_{\alpha\pm}^{*}(w) &=&
  \exp\left(\sum_{k=1}^{\infty}
    \frac{[2k]}{k}q^{\alpha k}w^k\beta_{k\pm}^{*}\right)
  \nonumber \\
  &=&
  \left(\frac{1-q^{2\alpha+1}w^2}{1-q^{2\alpha+3}w^2}\right)^{1/2}
  \frac{1}{1-q^{\alpha-1/2}r_\pm w},
\end{eqnarray}
\begin{eqnarray}
  &&
  \tau_{XX,\alpha}(\{\zeta_i\},\{w_a\}) =
  \exp\left(\sum_{k=1}^{\infty}\frac{[2k][k]}{k}q^{\alpha k}X_k X_k\right)
  \nonumber \\
  && =
  \prod_{i,i'=1}^{N}\prod_{a,a'\in A}
  \frac{(q^{\alpha+10}\zeta_i^2 \zeta_{i'}^{2};q^4)_\infty
    (1-q^{\alpha+4}\zeta_i^2 w_{a'})(1-q^{\alpha+4}\zeta_{i'}^{2}w_a)}
  {(q^{\alpha+8}\zeta_i^2 \zeta_{i'}^{2};q^4)_\infty
    (1-q^\alpha w_a w_{a'})(1-q^{\alpha+2}w_a w_{a'})}, \\
  &&
\label{eqn:form1}
  \tau_{XY,\alpha}(\{\zeta_i\},\{w_a\}) =
  \exp\left(\sum_{k=1}^{\infty}\frac{[2k][k]}{k}q^{\alpha k}X_k Y_k\right)
  \nonumber \\
  && =
  \prod_{i,i'=1}^{N}\prod_{a,a'\in A}
  \frac{(q^{\alpha+5}\zeta_i^2 \zeta_{i'}^{-2};q^4)_\infty
    (1-q^{\alpha+4}\zeta_i^2 w_{a'}^{-1})(1-q^{\alpha-2}\zeta_{i'}^{-2} w_a)}
  {(q^{\alpha+3}\zeta_i^2 \zeta_{i'}^{-2};q^4)_\infty
    (1-q^\alpha w_a w_{a'}^{-1})(1-q^{\alpha+2}w_a w_{a'}^{-1})}, \\
  &&
  \tau_{XX^*,\alpha}(\{\zeta_i\},\{w_a\};\{\zeta_j^*\},\{w_b^*\}) =
  \exp\left(\sum_{k=1}^{\infty}\frac{[2k][k]}{k}q^{\alpha k}X_k X_k\right)
  \nonumber \\
  && =
  \prod_{i,j=1}^{N}\prod_{a\in A \atop{b\in B}}
  \frac{(q^{\alpha+8}\zeta_i^2 \zeta_{j}^{*\: 2};q^4)_\infty
    (1-q^{\alpha+4}\zeta_i^2 w_b^*)(1-q^{\alpha+2}\zeta_{j}^{*\: 2}w_a)}
  {(q^{\alpha+6}\zeta_i^2 \zeta_{j}^{*\: 2};q^4)_\infty
    (1-q^\alpha w_a w_b^*)(1-q^{\alpha+2}w_a w_b^*)}, \\
  &&
  \tau_{XY^*,\alpha}(\{\zeta_i\},\{w_a\};\{\zeta_j^*\},\{w_b^*\}) =
  \exp\left(\sum_{k=1}^{\infty}\frac{[2k][k]}{k}q^{\alpha k}X_k Y_k^*\right)
  \nonumber \\
  && =
  \prod_{i,j=1}^{N}\prod_{a\in A\atop{b\in B}}
  \frac{(q^{\alpha+6}\zeta_i^2 \zeta_{j}^{*-2};q^4)_\infty
    (1-q^{\alpha+4}\zeta_i^2 w_{b}^{*-1})(1-q^\alpha \zeta_{j}^{*-2}w_a)}
  {(q^{\alpha+4}\zeta_i^2 \zeta_{j}^{*-2};q^4)_\infty
    (1-q^\alpha w_a w_{b}^{*-1})(1-q^{\alpha+2}w_a w_{b}^{*-1})}, \\
  &&
  \tau_{YX^*,\alpha}(\{\zeta_i\},\{w_a\};\{\zeta_j^*\},\{w_b^*\}) =
  \exp\left(\sum_{k=1}^{\infty}\frac{[2k][k]}{k}q^{\alpha k}Y_k X_k^*\right)
  \nonumber \\
  && =
  \prod_{i,j=1}^{N}\prod_{a\in A\atop{b\in B}}
  \frac{(q^{\alpha+2}\zeta_{i}^{-2}\zeta_{j}^{*\: 2};q^4)_\infty
    (1-q^{\alpha-2}\zeta_{i}^{-2}w_b^*)
    (1-q^{\alpha+2}\zeta_{j}^{*\: 2}w_{a}^{-1})}
  {(q^\alpha \zeta_{i}^{-2}\zeta_{j}^{*\: 2};q^4)_\infty
    (1-q^\alpha w_{a}^{-1}w_b^*)(1-q^{\alpha+2}w_{a}^{-1}w_b^*)}, \\
  &&
  \tau_{YY^*,\alpha}(\{\zeta_i\},\{w_a\};\{\zeta_j^*\},\{w_b^*\}) =
  \exp\left(\sum_{k=1}^{\infty}\frac{[2k][k]}{k}q^{\alpha k}Y_k Y_k^*\right)
  \nonumber \\
  && =
  \prod_{i,j=1}^{N}\prod_{a\in A\atop{b\in B}}
  \frac{(q^\alpha \zeta_{i}^{-2}\zeta_{j}^{*-2};q^4)_\infty
    (1-q^{\alpha-2}\zeta_{i}^{-2}w_{b}^{*-1})
    (1-q^{\alpha+2}\zeta_{j}^{*-2}w_{a}^{-1})}
  {(q^{\alpha-2}\zeta_{i}^{-2}\zeta_{j}^{*-2};q^4)_\infty
    (1-q^\alpha w_a w_{b}^{*-1})(1-q^{\alpha+2}w_a w_{b}^{*-1})}, \\
  &&
\label{eqn:form2}
  \tau_{X^* Y^*,\alpha}(\{\zeta_i^*\},\{w_b^*\}) =
  \exp\left(\sum_{k=1}^{\infty}\frac{[2k][k]}{k}q^{\alpha k}X_k^* Y_k^*\right)
  \nonumber \\
  && =
  \prod_{j,j'=1}^{N}\prod_{b,b'\in B}
  \frac{(q^{\alpha+4}\zeta_{j}^{*\: 2} \zeta_{j'}^{*-2};q^4)_\infty
    (1-q^{\alpha+2}\zeta_{j}^{*\: 2}w_{b'}^{*-1})
    (1-q^\alpha\zeta_{j'}^{*-2}w_{b}^{*})}
  {(q^{\alpha+2}\zeta_{j}^{*\: 2}\zeta_{j'}^{*-2};q^4)_\infty
    (1-q^\alpha w_{b}^* w_{b'}^{*-1})
    (1-q^{\alpha+2}w_{b}^* w_{b'}^{*-1})}, \\
  &&
  \tau_{Y^* Y^*,\alpha}(\{\zeta_i^*\},\{w_a^*\}) =
  \exp\left(\sum_{k=1}^{\infty}\frac{[2k][k]}{k}q^{\alpha k}Y_k^* Y_k^*\right)
  \nonumber \\
  && =
  \prod_{j,j'=1}^{N}\prod_{b,b'\in B}
  \frac{(q^{\alpha+2}\zeta_{j}^{*-2}\zeta_{j'}^{*-2};q^4)_\infty
    (1-q^\alpha \zeta_{j}^{*-2}w_{b'}^{*-1})
    (1-q^\alpha \zeta_{j'}^{*-2}w_{b}^{*-1})}
  {(q^\alpha \zeta_{j}^{*-2}\zeta_{j'}^{*-2};q^4)_\infty
    (1-q^\alpha w_{b}^{*-1}w_{b'}^{*-1})
    (1-q^{\alpha+2}w_{b}^{*-1}w_{b'}^{*-1})}.
\end{eqnarray}

\end{document}